\documentclass[floats,floatfix,amssymb,prd,twocolumn,superscriptaddress,nofootinbib]{revtex4-2}

\usepackage[english]{babel}
\usepackage[utf8]{inputenc}
\usepackage{amsmath}
\usepackage{mathbbol}
\usepackage{amssymb}
\usepackage{bbold}
\usepackage{graphicx,amsfonts}
\usepackage{epsfig}
\usepackage[colorlinks=true,
linkcolor=blue,
urlcolor=blue,
citecolor=blue]{hyperref}
\usepackage{bm}
\usepackage{mathrsfs}
\usepackage{mathtools}
\usepackage{enumerate}
\usepackage{amsthm}
\usepackage{bbm}
\usepackage{comment}
\usepackage{physics}
\usepackage{url}
\usepackage{upgreek}
\usepackage[svgnames]{xcolor}
\usepackage[title]{appendix}

\begin{document}

%%%%%%%%%%%%%%%%%%%%%%%%%%%%%%%%%%%%%%%%%%%%%%%
\title{Superradiance of charged black holes embedded in dark matter halos}
%%%%%%%%%%%%%%%%%%%%%%%%%%%%%%%%%%%%%%%%%%%%%%%
\author{Alessandro Mollicone}
\affiliation{Dipartimento di Fisica, Sapienza Università di Roma \& INFN, Sezione di Roma, Piazzale Aldo Moro 5, 00185, Roma, Italy}

\author{Kyriakos Destounis}
\affiliation{Dipartimento di Fisica, Sapienza Università di Roma \& INFN, Sezione di Roma, Piazzale Aldo Moro 5, 00185, Roma, Italy}
%%%%%%%%%%%%%%%%%%%%%%%%%%%%%%%%%%%%%%%%%%%%%%%

%%%%%%%%%%%%%%%%
\begin{abstract}
Astrophysical environments are ubiquitous in the Universe; from accretion disks around black holes to galactic dark matter halos, distributions of astrophysical material veil the vast majority of Cosmos. Including environmental effects in strong-field gravity and astrophysics is, therefore, a rather tantalizing task in the quest for novel gravitational-wave phenomena. Here, we examine how environments affect the high-energy process of superradiance. In particular, we study the amplification of charged scalar waves under the expense of the electrostatic energy contained in a charged black hole that is embedded in an observationally-motivated, and qualitatively generic, dark matter halo. We find that the superradiant amplification of massless charged scalar fields scattering off environmentally-enriched charged black holes can be equally efficient  to those occurring in vacuum charged black holes. This occurs due to the fact that the sole interplay between the scalar wave and the black hole is the electromagnetic interaction. The addition of mass on the charged scalar waves leads to a rapid suppression of superradiant amplification. This transpires to a great extent due to the `friction' that the mass introduces to the black hole potential. Nevertheless, for sufficiently large scalar masses, the amplification factors can be also subdominantly affected by the compactness of the halo. This occurs because the gravitational interaction between the dense halo and the wave's mass grows, thus further suppressing the superradiant amplification.
\end{abstract}
%%%%%%%%%%%%%%%%

\maketitle

%%%%%%%%%%%%%%%%%%%%%%
\section{Introduction}
%%%%%%%%%%%%%%%%%%%%%%

The apparent success of gravitational-wave (GW) astronomy has led to an excess of new fields of research that regard, not only the phenomenology of black holes (BHs) and compact objects, but also the astrophysical environments around them, and even the structure of galaxies that harbor supermassive BHs in their core. To that end, a well-known discrepancy exists between the experimental rotation curves of galaxies, and the curves derived only by taking into account the baryonic matter observed in a galaxy \cite{Roberts:1975,Mareki:1997}. A new particle has been proposed for this purpose, i.e. dark matter \cite{Profumo:2019ujg}, that forms clumps and halos within which galaxies are born \cite{Cole:1991,Primack:2009}. Such halos are currently the main postulated solution to account for the rotation curve variance between theory and observations in galaxies.

Dark matter, together with dark energy, pose one of the most extraordinary problems to the current cosmological standard model \cite{Oks:2021hef}. Even though dark matter particles have not yet been directly detected and are absent from the standard model of particle physics \cite{Hooper:2009zm}, their indirect role has been established, through observations and numerical simulations, in cosmology \cite{Arbey:2021gdg}, structure formation \cite{Blanton:2006tw,Bower:2005vb}, galactic dynamics \cite{Wechsler:2018pic,Behroozi:2019kql} and recently in astrophysical environments around compact objects \cite{Cardoso:2021wlq,Spieksma:2024voy}. Dark matter overdensities (spikes) close to BH horizons and galactic dark matter halos have enabled us to examine its large-scale phenomenology and potential particle attributes in a general-relativistic perspective. The field of environmental effects around supermassive, or otherwise, BHs can, therefore, shed more light into the nature of dark matter \cite{Frampton:2010sw,Ivanov:1994pa,Buckley:2017ijx,Villanueva-Domingo:2021spv} and uncover a surfeit of aspects of astrophysical environments \cite{Barausse:2014tra,Cardoso:2019upw,Cardoso:2022whc,Cardoso:2021wlq,Jaramillo:2020tuu,Jaramillo:2021tmt,Jaramillo:2022kuv,Destounis:2020kss,Destounis:2021mqv,Destounis:2021lum,Cheung:2021bol,Berti:2022xfj,Boyanov:2022ark,Courty:2023rxk,Destounis:2023nmb,Boyanov:2023qqf,Cole:2022yzw,Annulli:2020lyc,Duque:2023seg,Eleni:2024fgs,Strateny:2023edo}.

To this day, observations lead the race in understanding the role of dark matter in galaxies and, more generally, our Cosmos \cite{Persic:1995ru,Ludwig:2021kea,Sands:2024vvd}. Various halo distributions have been proposed that provide a plethora of density profiles that are appropriate for dark matter overdensities, galactic bulges, spiral, as well as elliptical galaxies. Even though these dark matter distributions are designed from observations and structure formation simulations, they do not choose a particular dark matter particle candidate but focus more on the large-scale effects of concentrated or sparse dark matter \cite{Bertone:2004pz}. Some examples of dark matter distributions are the Hernquist profile \cite{Hernquist:1990}, that provides an analytical model for spherical galaxies and bulges \cite{Cardoso:2021wlq,Cardoso:2022whc,Destounis:2022obl}, the Navarro-Frenk-White profile \cite{Navarro:1995iw}, that is fitted to dark matter halos in N-body simulations and the Einasto model \cite{Einasto:1968}, that describes how the density of a spherical stellar system varies with the distance from its center, among others.

In Ref. \cite{Cardoso:2021wlq}, the first exact, fully-relativistic solution to the field equations was obtained that describes a static, spherically-symmetric BH surrounded by a Hernquist dark matter distribution. The methodology used to construct the aforementioned `galactic' BH solution has been extended in various directions \cite{Jusufi:2019,Cardoso:2022whc,Konoplya:2022hbl,Jusufi:2022jxu,Destounis:2022obl,Feng:2022evy,Figueiredo:2023gas,Shen:2023erj,Datta:2023zmd,Speeney:2024mas,Heydari-Fard:2024wgu} and till now, it has been proven useful in studying environmental effects on GW generation and propagation. An interesting generalization of the spacetime in Ref. \cite{Cardoso:2021wlq} appeared recently \cite{Stelea:2023yqo} that introduced a charge to the central BH while keeping the dark matter halo intact (of the Hernquist type). This is a direct generalization of the Ehlers-Harrison transformations in the Ernst formalism \cite{Ernst1,Ernst2,Ehlers,Harrison} utilized in order to construct in an analytical form the charged version of Ref. \cite{Cardoso:2021wlq}.

Since a generalization to include rotation in Ref. \cite{Cardoso:2021wlq} has not yet been achieved for any dark matter distribution, a good `toy model' to study effects that take place in rotating BHs at the center of a galaxy is a charged realization of the configuration. Indeed, rotating and charged vacuum BHs possess a strikingly similar causal structure, thus making charged BHs ideal testbeds for high-energy phenomena\footnote{Other phenomenology has also been studied, such as quasinormal modes \cite{Cardoso:2021wlq,Cardoso:2022whc,Konoplya:2021ube,Daghigh:2023ixh}, optical appearance \cite{Daghigh:2023ixh,Macedo:2024qky,Xavier:2023exm}, extreme-mass-ratio binary fluxes \cite{Cardoso:2022whc,Figueiredo:2023gas,Rahman:2023sof,Speeney:2024mas} and greybody factors \cite{Konoplya:2021ube,Rosato:2024arw}, among others.}, such as the process of energy extraction from a dissipative system and BH superradiance \cite{Brito:2015oca}.

Superradiance occurs in a variety of settings and a plethora of spacetimes in General Relativity (GR), as long as the BH includes rotation or charge. In particular, charged BHs exhibit superradiance due to the electrostatic interaction between the electric charge of the BH and the electric charge of the scalar field that is impinging the BH \cite{DiMenza:2014vpa,Myung:2022dpp,Destounis:2022rpk}. The charged scalar field is partially reflected when encountering the BH light-ring and extracts electromagnetic energy from the generalized ergoregion to the wave zone \cite{Denardo:1973pyo,Baez:2024lhn}\footnote{A similar analysis can be used to study superradiant instabilities of magnetically-charged BHs \cite{Brito:2014nja}, as well as magnetic rotating primordial BHs in order to assess the existence of magnetic monopoles \cite{Pereniguez:2024fkn}.}. Superradiance is a dynamical process and can be quenched after a particular amount of energy is extracted from the BH \cite{Brito:2015oca}. Nevertheless, in extreme scenarios, superradiance can amplify the scattered charged scalar wave if its oscillation frequency belongs to the frequency range where superradiance occurs. In this case, the scalar wave's amplitude grows in time, due to superradiance, and leads to a superradiant instability \cite{Cardoso:2018nvb,Destounis:2019hca,Mascher:2022pku,Zhu:2014sya,Konoplya:2014lha} and the eventual explosion of the `BH bomb' \cite{Cardoso:2004nk,Hod:2009cp,Cardoso:2013krh,DiMenza:2019zli}. 

BH superradiance has shown to be valuable in several aspects of gravity and holds quite a promising discovery potential. It is used in order to better understand the formation of ultralight bosonic clouds around BHs \cite{Berti:2019wnn}, as well as their particle nature \cite{Chan:2022dkt,Sun:2019mqb,Isi:2018pzk}, that may shed light into beyond the standard model physics, such as dilatons, axions and dark matter \cite{Hannuksela:2018izj}. More recent studies that regard BH/dark matter configurations have shown that superradiance is suppressed in the presence of dark matter halos \cite{Liu:2022ygf,Liu:2024qso}. Yet, the dark matter clouds were consisted of cold, scalar field or perfect fluid dark matter. 

In this analysis, we study the phenomenon of superradiance by scattering massless and massive charged monochromatic scalar fields off the spacetime consisting of a charged BH surrounded by a Hernquist-type dark matter halo. It is worthy to note that the dark matter content in the spacetimes \cite{Cardoso:2021wlq,Stelea:2023yqo} does not assume a specific particle type for the galactic halo, as performed in Refs. \cite{Liu:2022ygf,Liu:2024qso}, but rather uses an anisotropic fluid with energy density and tangential pressure resulting from galaxy formation simulations and experimental observations. We find that massless charged scalar waves that scatter off the aforementioned BH spacetime lead to similar amplification factors as in vacuum Reissner-Nordstr\"om (RN) BHs \cite{Reissner,Weyl,Nordstrom}. This is due to the fact that the only interaction of the wave's charge occurs with the electrostatic charge of the BH, since the halo is neutral. Massive charged scalar waves have a suppressive influence on superradiant amplification. This practically takes place mostly due to the `friction-like' role that the scalar mass has in the effective potential of massive charged scalar perturbations, though when the halo becomes more dense and compact a sudominant gravitational effect occurs between the scalar mass and the halo that further suppresses superradiance. In what follows, we adopt the geometrized units so that $G=c=1$.

%%%%%%%%%%%%%%%%%%%%%%%%%%%%%%%%%%%%%%%%%%%%%%%%%%%%%
\section{Black holes surrounded by dark matter halos}
%%%%%%%%%%%%%%%%%%%%%%%%%%%%%%%%%%%%%%%%%%%%%%%%%%%%%

%%%%%%%%%%%%%%%%%%%%%%%%%%%%%%%%%%%%%%%%%%%%%%%%%%%
\subsection{Neutral BH surrounded by Hernquist dark matter}
%%%%%%%%%%%%%%%%%%%%%%%%%%%%%%%%%%%%%%%%%%%%%%%%%%%

The first fully-relativistic solution of GR that describes an exact, static and spherically-symmetric BH enclosed in a dark matter halo profile of the Hernquist-type was obtained in \cite{Cardoso:2021wlq}. The environment is `inserted' in the BH configuration by the Einstein cluster mechanism which results in a stress-energy tensor that describes an anisotropic fluid with tangential and no radial pressure, i.e. $T^{\mu}_{\nu} = \text{diag}(-\rho,0,P_t,P_t)$. The energy density of the anisotropic fluid has the form \cite{Hernquist:1990}
\begin{equation}\label{Hernquist_profile}
    \rho(r) = \frac{M a_0}{2 \pi r (r+a_0)^3},
\end{equation}
where $M$ and $a_0$ are the dark matter halo's mass and length scale, respectively. According to observations, Eq. \eqref{Hernquist_profile} provides a description of the observed density in spiral and elliptical galaxies, as well as galactic bulges \cite{Hernquist:1990}. The line element of spacetime is  
\begin{equation}\label{eq:uncharged_metric}
    ds^2 = - f(r) dt^2 + \frac{dr^2}{1-\frac{2m(r)}{r}} + r^2(d \theta^2 + \sin^2 \theta d \varphi^2),
\end{equation}
where $m(r)$, i.e. the mass function of spacetime, is chosen to comply with observations and has the form
\begin{equation}\label{mass_function_m(r)}
    m(r) = M_{\textrm{BH}} + \frac{Mr^2}{(a_0 + r)^2}\Bigg( 1 - \frac{2 M_{\textrm{BH}}}{r} \Bigg)^2,
\end{equation}
where $M_{\textrm{BH}}$ is the BH mass. At small distances, Eq. \eqref{mass_function_m(r)} describes a static BH of mass $M_{\textrm{BH}}$, while at large distances it describes the Hernquist dark matter profile (\ref{Hernquist_profile}). The lapse function $f(r)$ can be obtained from $m(r)$ by imposing that the $(r,r)$ component of Einstein's equations vanishes (i.e. zero radial pressure). This leads to the exact solution \cite{Cardoso:2021wlq} with
\begin{equation}\label{eq:f_Cardoso}
    f(r) = \Bigg( 1 - \frac{2 M_{\textrm{BH}}}{r} \Bigg) e^{\Upsilon},
\end{equation}
where 
\begin{equation}
\begin{split}
    &\Upsilon = - \pi \sqrt{\frac{M}{\xi}} + 2 \sqrt{\frac{M}{\xi}}\arctan \frac{r + a_0 - M}{\sqrt{M \xi}},\\
    &\xi = 2 a_0 - M + 4M_{\textrm{BH}}.
\end{split}    
\end{equation}
The BH possesses an event horizon at $r = 2M_{\textrm{BH}}$ and a curvature singularity at $r=0$. To ensure the absence of external curvature singularities and $\xi>0$, the scales of spacetimes satisfy $M_{\textrm{BH}} \ll M \ll a_0$.

%%%%%%%%%%%%%%%%%%%%%%%%%%%%%%%%%%%%%%%%%%%%%%%%%%%
\subsection{Charged BH surrounded by Hernquist dark matter}
%%%%%%%%%%%%%%%%%%%%%%%%%%%%%%%%%%%%%%%%%%%%%%%%%%%

An exact, analytic solution that describes a charged BH immersed in a Hernquist dark matter halo was recently obtained in Ref. \cite{Stelea:2023yqo}. This solution is obtained by the means of a charging technique that is described in \cite{2018arXiv181002235S}. Through a solution-generating scheme, which is a direct generalization of certain Ehlers-Harrison transformations in the Ernst formalism, we begin from any static, axially-symmetric geometry, which in general is sourced by an anisotropic fluid described by a non-diagonal stress-energy tensor, and end up with a charged (or magnetized) counterpart. The resulting electrically charged version of Eq. (\ref{eq:uncharged_metric}) has a line element of the form \cite{Stelea:2023yqo}
\begin{equation}\label{eq:Stelea_line_element}
    ds^2 = - \frac{f(r)}{\Lambda^2(r)}dt^2 + \Lambda^2(r) \left( \frac{dr^2}{b(r)} + r^2 d\Omega^2_2 \right),
\end{equation}
where $d\Omega^2_2$ is the line element of the unit 2-sphere, $f(r)$ is the lapse function from Eq. \eqref{eq:f_Cardoso}, while $b(r) = 1 - 2m(r)/r$. The function $\Lambda(r)$ is defined as
\begin{equation}\label{Lambda_Stelea_et_al}
    \Lambda(r) = \frac{1 - U^2 f(r)}{1 - U^2},\ \ \ \ \ \ U \in [0,1),
\end{equation}
while the electromagnetic four-potential is defined as 
\begin{equation}\label{eq:EM_4_potential}
    A_{\mu} = \left(\frac{U f(r)}{\Lambda(r)},0,0,0 \right).
\end{equation}
Then, the system of Einstein-Maxwell-fluid equations (see Eq. (12) in \cite{Stelea:2023yqo}) can be satisfied if we consider the stress-energy tensor of the fluid as
\begin{equation}
    T_{\mu\nu}=(\rho+\rho_e+P_t)u_\mu u_\nu+P_t(g_{\mu\nu}-\chi_\mu\chi_\nu),
\end{equation}
while the stress-energy of the electromagnetic has the typical form
\begin{equation}
    T^{em}_{\mu\nu}=\frac{1}{4\pi}\left(F_{\mu\gamma}F_{\nu}^{\gamma}-\frac{1}{4}g_{\mu\nu}F_{\gamma\delta}F^{\gamma\delta}\right).
\end{equation}
Here, $u^\mu=\Lambda(r)\delta_t^\mu/f^{-1/2}(r)$ is the four-velocity of the fluid, normalized so that $u_\mu u^\mu=-1$, while $\chi^\mu=b(r)\delta_r^\mu/\Lambda(r)$ is the unit vector in the radial direction. Consequently, $\rho\rightarrow \rho/\Lambda^2$ and $P_t\rightarrow P_t/\Lambda^2$, while the effective stress-energy tensor generated by the source term of the Maxwell-fluid Lagrangian is
\begin{equation}
\rho_e=\frac{2\left(\rho+2P_t\right)}{1-U^2}\frac{U^2 f(r)}{\Lambda(r)}.
\end{equation}

When $U=0$, then $\Lambda(r)=1$, and Eq. (\ref{eq:Stelea_line_element}) reduces to the neutral solution of Eq. (\ref{eq:uncharged_metric}). The RN limit can be obtained by setting $M = 0$, $a_0=0$ and performing the coordinate transformation $R=\Lambda r$, which is equivalent to \cite{Stelea:2023yqo}
\begin{equation}\label{eq:coordinate_transformation_RN_limit}
    r = R - \frac{2 M_{\textrm{BH}} U^2}{1 - U^2}.
\end{equation}
In the RN limit, 
\begin{equation}\label{eq:ADM_and_charge}
    M_{\text{ADM}}=\frac{1+U^2}{1-U^2}M_{\textrm{BH}},\,\,\,\,\,\,\,\,
    Q_{\infty}=\frac{2U}{1-U^2}M_{\textrm{BH}},
\end{equation}
are the ADM mass and the charge associated with the electric field measured at infinity, respectively. 

In the asymptotic region ($r\rightarrow \infty$), $\Lambda\rightarrow 1$ and the geometry \eqref{eq:Stelea_line_element} is asymptotically flat. There, the ADM mass of the configuration is
\begin{equation}\label{eq:ADM_mass}
    M_\textrm{ADM}=\frac{1+U^2}{1-U^2}\left(M_\textrm{BH}+M\right).
\end{equation}
The ADM mass contains a contribution from both the BH mass $M_\textrm{BH}$ and the dark matter halo, $M$. One can also expect that the asymptotic electric charge of the charged BH surrounded by a Hernquist dark matter
halo \eqref{eq:Stelea_line_element}, will be given by
\begin{equation}\label{Q_inf}
    Q_\infty=\frac{2U \left(M_\textrm{BH}+M\right)}{1-U^2}.
\end{equation}
In turn, the radius $r=2M_\textrm{BH}$ corresponds to the BH event horizon (we will return to this crucial aspect of the BH in study in the next paragraph). Thus, the near-horizon BH charge (namely, in the limit $r\rightarrow 2M_\textrm{BH}$), after a similar analysis, takes the form
\begin{equation}\label{eq:BH_charge}
    Q_\textrm{BH}=\frac{2 M_\textrm{BH} U}{1-U^2}e^\Upsilon,
\end{equation}
where at this limit, $e^\Upsilon\rightarrow 1-2M/a_0$.
Therefore, in the charged configuration, the extremal limit is approached as $Q_\infty/M_\textrm{ADM}=2U/(1+U^2)\rightarrow 1$, i.e. when $U\rightarrow 1$. Thus, these configurations have an extremal limit beyond which there are no BH solutions \cite{Stelea:2023yqo}, as in RN spacetimes ($Q\rightarrow M$).

Despite Eq. \eqref{eq:Stelea_line_element} being the line element of a charged solution to the Einstein-Maxwell-fluid equations, the metric (\ref{eq:Stelea_line_element}) exhibits only one horizon at $r=2M_{\textrm{BH}}$, as mentioned above. This unique property arises due to the charging technique employed \cite{Stelea:2018cgm,Stelea:2018elx,Stelea:2018shx,Stelea:2023yqo}. Specifically, by working with a generalization of GR where Harrison-like transformations of the Ernst electromagnetic potential become vector equations, it has been shown that in the simplest case of scalar equations the Cauchy horizon is pushed towards the origin, that further result to the introduction of electric and magnetic charges to a stationary, axially-symmetric spacetime \cite{Galtsov:1996qko}.  Indeed, in the RN limit, in order to obtain the standard form of the RN metric (and include explicitly the BH charge in the metric), Ref. \cite{Stelea:2023yqo} used the coordinate transformation (\ref{eq:coordinate_transformation_RN_limit}), which for $r = 0$ reduces to $R = \frac{2M_\textrm{BH}U^2}{1-U^2}$. It is then straightforward to show that
\begin{equation}
    R_- = M_\textrm{ADM} - \sqrt{M_\textrm{ADM}^2 - Q_{\infty}^2} = \frac{2M_\textrm{BH}U^2}{1-U^2},
\end{equation}
where $R_-$ is the null Cauchy horizon in Schwarschild-like coordinates.

We note that in what follows, we will use $M_\textrm{BH}=1$ (if not stated otherwise) for the central BH, as we would perform in a vacuum RN BH, though the total ADM mass of the system will be controlled by the quantity in Eq. \eqref{Q_inf}. This will affect the superradiant amplification results by a factor of $M_\textrm{ADM}$ which is not equal to unity but rather a much larger quantity, due to the halo mass. In turn, similar corresponding amplification factors in vacuum RN are obtained after normalizing the extremal limit as ${Q_\infty}
/M_\textrm{ADM}$ and not as ${Q_\infty}
/M_\textrm{BH}$, meaning that we will explore superradiance for some cases in the RN limit where $M_\textrm{BH}>1$, thus $M_\textrm{BH}>Q_\infty>1$ to ensure the existence of a BH horizon.

%%%%%%%%%%%%%%%%%%%%%%%%%%%%%%%%%%%%%%
\section{Charged scalar perturbations}
%%%%%%%%%%%%%%%%%%%%%%%%%%%%%%%%%%%%%%

In this section we analyze how a massive, charged scalar field $\phi$ propagates on the fixed background \eqref{eq:Stelea_line_element}. Specifically, We are interested in a scattering scenario in order to investigate the existence of superradiance on the aforementioned background geometry. We assume that the amplitude of the scalar field is sufficiently small, therefore it does not backreact to the spacetime metric.
In other words, we treat the field’s contribution to the stress-energy tensor as a second-order effect. The equation governing the propagation of a monochromatic massive charged scalar waves on a curved spacetime, i.e. the charged Klein-Gordon equation, has the form
\begin{equation}\label{eq:Klein-Gordon_eq}
    (D_{\mu}D^{\mu} - \mu^2)\phi = 0,
\end{equation}
where $D_{\mu} = \nabla_{\mu} - iqA_{\mu}$ is the 'gauge' covariant derivative, $\nabla_{\mu}$ is the spacetime covariant derivative, $q$ is the scalar charge, $A_\mu$ is the electrostatic potential given by Eq. (\ref{eq:EM_4_potential}) and $\mu^2$ represents the scalar mass. In general, $\phi = \phi(t,r,\theta,\varphi)$. Due to spherically symmetry, we can separate the radial and temporal parts from the angular part of the scalar field by choosing the following ansatz:
\begin{equation}\label{eq:ansatz}
    \phi(t,r,\theta,\varphi) = e^{-i \omega t} \sum_{\ell,m}\frac{\Phi_{\ell m}(r)}{r} Y_{\ell m} (\theta, \varphi),
\end{equation}
where $\ell$ and $m$ are the angular index and the azimuthal number, respectively. Then, by substituting Eq. \eqref{eq:ansatz} in \eqref{eq:Klein-Gordon_eq}, the Klein-Gordon equation can be reduced to a second-order ordinary differential equation for the radial part of $\phi$ as
\begin{align}\nonumber
 \frac{d^2 \Phi}{d r_{*}^2} &+ \left( \frac{2 \Lambda'(r) \sqrt{f(r)b(r)}}{\Lambda^3(r)} \right)\frac{d \Phi}{d r_{*}} \\\label{eq:Schrodinger_Stelea}
 &+ \left(\omega^2 + 2\omega\frac{f(r) qU}{\Lambda(r)} - V_{\ell}(r) \right)\Phi = 0,
\end{align}
where the subscripts $(\ell,m)$ have been omitted for simplicity, while the tortoise coordinate $r_*$ is defined as 
\begin{equation}\label{eq:tortoise_Stelea}
    dr_{*} = \frac{\Lambda^2(r)}{\sqrt{f(r)b(r)}} dr.
\end{equation}
The effective potential $V_\ell(r)$ in Eq. (\ref{eq:Schrodinger_Stelea}) is given by
\begin{align}\nonumber
   V_{\ell}(r) &= \frac{f(r)}{\Lambda^2(r)} \left( \frac{1}{2 r \Lambda^2(r)} \left[\frac{b(r) f'(r)}{f(r)}+b'(r)\right] \right.\\\label{eq:effective_potential}
   & \left.+ \frac{\ell(\ell+1)}{r^2 \Lambda^2(r)} + \mu^2 \right)-\left(\frac{f(r) qU}{\Lambda(r)}\right)^2.
\end{align}
We note that Eq. (\ref{eq:Schrodinger_Stelea}) does not have the typical Schr\"odinger-like form. In fact, the second term, that represents a first order derivative of the radial part of the field with respect to the tortoise coordinate, encodes a different dynamical evolution of scalar perturbations than that of simple Schr\"odinger-like equations. This term's occurrence arises due to the presence of  $\Lambda(r)$ in the differential equation that defines the tortoise coordinate, i.e. Eq. (\ref{eq:tortoise_Stelea}). Interestingly, since $\Lambda(r)$ depends on the lapse function $f(r)$ and $b(r)$, that include both contributions from the halo, it incorporate the environmental effects that modify the BH potential. From this point on we will refer to $qUM_\textrm{BH}$ as the `charge coupling' between the scalar field and the BH. Further justification of the role of $qUM_\textrm{BH}$ as the charge coupling will be given later (see section IV).

\begin{figure*}[t]
     \includegraphics[width=\textwidth]{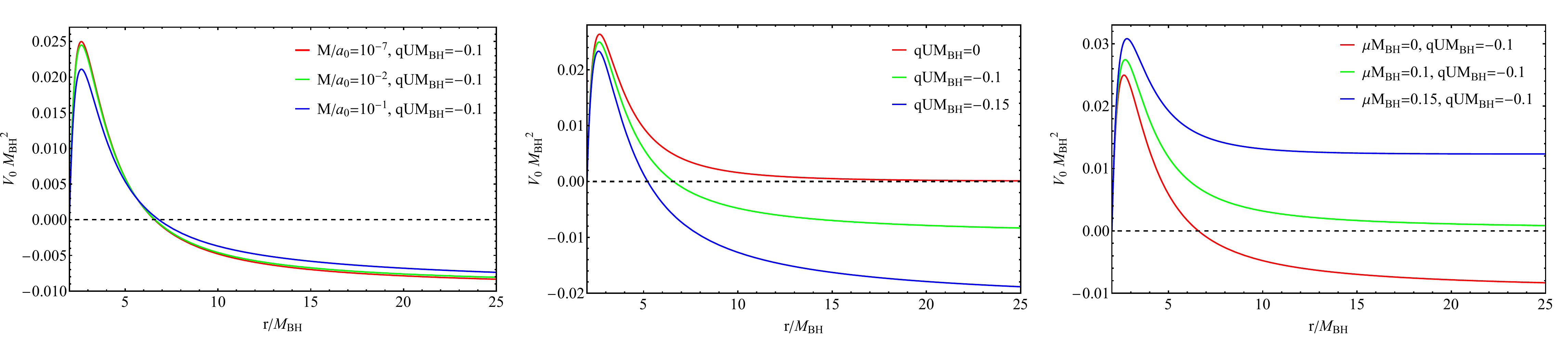}
     \caption{Effective potential \eqref{eq:effective_potential} for $\ell=0$ and a variety of choices for the parameters $M/a_0,\,qU M_\textrm{BH}$ and $\mu M_\textrm{BH}$. \emph{Left:} Effective potential with $\mu M_\textrm{BH}=0,\,qU M_\textrm{BH}=-0.1$ and varying compactness $M/a_0$. \emph{Middle:} Same for $\mu M_\textrm{BH}=0,\,M/a_0=10^{-4}$ and varying charge coupling $qUM_\textrm{BH}$. \emph{Right:} Same for $M/a_0=10^{-4},\,qU M_\textrm{BH}=-0.1$ and varying scalar mass $\mu M_\textrm{BH}$.}\label{fig:potentials}
\end{figure*}

In Fig. \ref{fig:potentials} we depict three different cases of the BH potential where we vary the compactness $M/a_0$, the charge coupling $qUM_\textrm{BH}$ and the scalar field mass $\mu M_\textrm{BH}$. It is obvious that the compactness does not affect significantly the effective potential (as it also occurs in the neutral case \eqref{eq:uncharged_metric}), therefore we should not expect a significant alteration on the amplification factors when the compactness is not strongly varied. On the other hand, the charge coupling reduces the potential to negative asymptotic values, while the scalar mass counterbalances the aforementioned effect by enhancing the effective potential to asymptotic values that are positive, thus partially cancelling the negative potential well which drives superradiance. We will see in later sections how the counterbalance mechanism between mass interaction and charge coupling is taking place, through amplification factor calculations.

%%%%%%%%%%%%%%%%%%%%%%%%%%%%%%%%%%%%%%
\section{Scattering and superradiance}
%%%%%%%%%%%%%%%%%%%%%%%%%%%%%%%%%%%%%%

To perform a scattering experiment involving massless and massive charged scalar waves impinging the geometry \eqref{eq:Stelea_line_element}, we solve numerically Eq. \eqref{eq:Schrodinger_Stelea}, by imposing the following boundary conditions
\begin{equation}\label{eq:asym_behav_Stelea}
    \Phi \sim
    \begin{cases}
      \mathcal{T}e^{-i \omega r_{*}},\,\,\,\,\,\,\,\,\,\,\,\,\,\,\,\,\qquad\qquad r_{*} \rightarrow -\infty\\
      \mathcal{I}e^{-i k_{\infty} r_{*}} + \mathcal{R} e^{i k_{\infty} r_{*}},\quad\,\, r_{*} \rightarrow+ \infty
    \end{cases} 
\end{equation}
where $\mathcal{I},\,\mathcal{R}$ and $\mathcal{T}$ are the incident, reflection and transmission coefficients, respectively. The scattered massive charged wave has $k_\infty = \pm \sqrt{(\omega + qU)^2 - \mu^2}$. The conservation of the Wronskian at the boundaries yields the relation 
\begin{equation}\label{eq:superradiant_R_Stelea_massive}
    |\mathcal{R}|^2 = |\mathcal{I}|^2 - \frac{\omega}{k_{\infty}} |\mathcal{T}|^2 = |\mathcal{I}|^2 + \frac{\omega}{\sqrt{(\omega + qU)^2 - \mu^2}} |\mathcal{T}|^2, 
\end{equation}
where we have chosen $k_\infty = - \sqrt{(\omega + qU)^2 - \mu^2}$, and assumed  $\omega > 0$ in order for the prefactor of $|\mathcal{T}|^2$ to be positive\footnote{An alternative, though unphysical, way to have a positive prefactor is to choose $k_\infty =\sqrt{(\omega + qU)^2 - \mu^2}$ but this requires for the wave frequency to satisfy the unbounded relation $\omega<0$ for superradiance to take place.}. With the above assumptions, Eq. \eqref{eq:superradiant_R_Stelea_massive} can lead to reflected waves at infinity that have larger amplitude ($\mathcal{R}$) than their initial amplitude ($\mathcal{I}$) and these waves with frequency $\omega$ are superradiant. Of course, for this to occur, $k_\infty$ must be a real number, therefore the relation $(\omega + qU)^2 - \mu^2>0$ should be satisfied. In the simplest case, where $qU\rightarrow 0$, we obtain $\mu<\omega$ which serves as a lower bound for waves to propagate at infinity, otherwise we just have decaying functions (see Section 4.3.1 in Ref. \cite{Brito:2015oca}). In the general case, Eq. \eqref{eq:superradiant_R_Stelea_massive} leads to the superradiance condition
\begin{equation}\label{eq:massive_superradiant_condition}
    \mu < \omega < -\mu + |qU|,
\end{equation}
where we will assume appropriate $qU$, in order for the relation $|qU| > 2 \mu$ to be satisfied. Finally, we note that Eq. (\ref{eq:massive_superradiant_condition}) reduces to 
\begin{equation}\label{eq:Stelea_superradiant_condition}
    0 < \omega < -qU,
\end{equation}
when $\mu = 0$. Here, we require $-qU > 0$, hence, $q<0$, since $U \in (0,1]$. However, we can rescale $U$ to be within the range $(-1,0]$, without modifying the metric or any perturbation equations relevant to superradiance, and use a positive scalar charge $q>0$. Comparing Eq. (\ref{eq:Stelea_superradiant_condition}) with the superradiant condition of massles waves scattering off vacuum charged BHs (see for instance Eq. (4.54) of \cite{Brito:2015oca}), we conclude that the strength of the electromagnetic coupling is controlled by the charge of the scalar $q$ and $U$, that is connected with the charge of the BH at the center of the configuration \eqref{eq:Stelea_line_element}. This is the reasoning behind naming $qUM_{\textrm{BH}}$ as the dimensionless charge coupling. 

%%%%%%%%%%%%%%%%%%%%%%%%%%%%%%%
\section{Amplification factors}
%%%%%%%%%%%%%%%%%%%%%%%%%%%%%%%

%%%%%%%%%%%%%%%%%%%%%%%%%%%%%%%%
\subsection{Numerical technique}
%%%%%%%%%%%%%%%%%%%%%%%%%%%%%%%%

We consider scattering experiments of monochromatic charged, massless and massive, scalar fields, with varying frequency $\omega$. We define the amplification factors as
\begin{equation}
    \mathcal{Z}_\ell = \frac{|\mathcal{R}|^2}{|\mathcal{I}|^2} -1,
\end{equation}
where we remind the reader that $\mathcal{I}$ and $\mathcal{R}$ are the incident wave amplitude and the reflection coefficient, respectively. We calculate them for a variety of frequencies $\omega$ by integrating Eq. (\ref{eq:Schrodinger_Stelea}) numerically, with boundary conditions as in Eq. (\ref{eq:asym_behav_Stelea}). The numerical calculation involves expanding the solutions of Eq. (\ref{eq:Schrodinger_Stelea}) at the event horizon and at infinity to arbitrary order (till we reach numerical convergence) and then match the two asymptotic solutions at an intermediate point by imposing regularity of the corresponding eigenfunctions and their derivatives.

%%%%%%%%%%%%%%%%%%%%%%%%%%%%%%%%%%
\subsection{Numerical convergence}
%%%%%%%%%%%%%%%%%%%%%%%%%%%%%%%%%%

\begin{figure}[t]
     \includegraphics[width=0.45\textwidth]{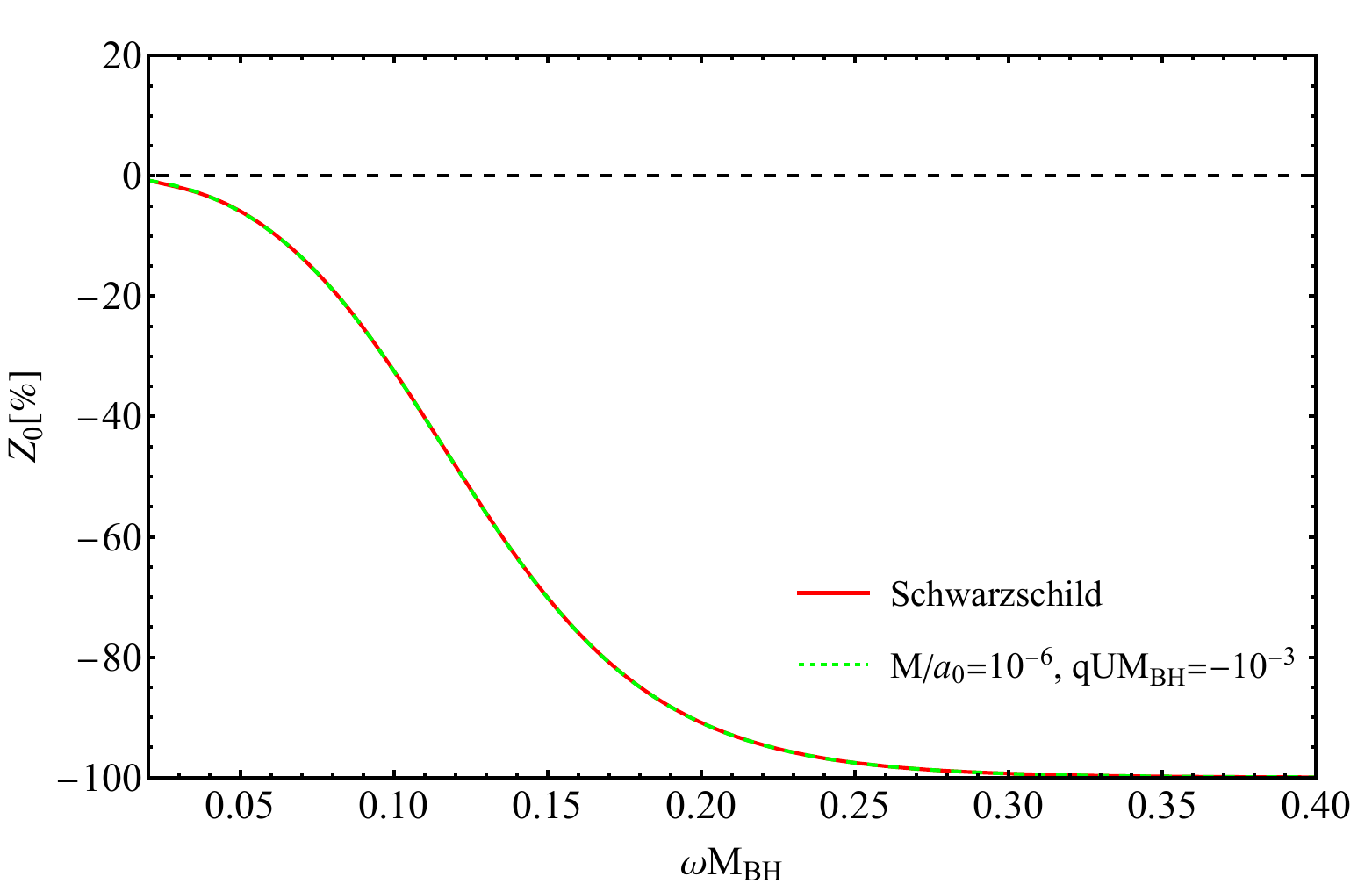}
     \caption{Amplification factors of a charged scalar field scattering off a Schwarzschild BH (shown with the red line) with $\ell=0$. Amplification factors at the Schwarzschild limit (shown with the dashed green line) of the geometry \eqref{eq:Stelea_line_element}, where $M/a_0=10^{-6}$, $qUM_\textrm{BH}=10^{-3}$ and $\ell=0$.}
     \label{fig:test_Schwarzschild}
\end{figure}

In Fig. \ref{fig:test_Schwarzschild} we perform a convergence test of our numerical code in order to understand if it satisfies appropriate limits, such as the Schwarzschild BH \cite{Schwarzschild} limit where superradiance does not take place. We use an already well-studied convergent numerical code that calculates the amplification factors for RN BHs, when charged scalars are scattered on them. By shutting down the charge of the scalar field or the BH charge we achieve the Schwarzschild limit and we can calculate the amplification factors on Schwarzschild spacetime. As expected from the theory, Schwarzschild BHs do not superradiate. This is in agreement with the numerical results we obtained (shown in red in Fig. \ref{fig:test_Schwarzschild}). Now, we use a novel numerical code that has been designed for the spacetime \eqref{eq:Stelea_line_element} and choose a small compactness, $M/a_0=10^{-6}$, while we choose a menial charge coupling, i.e. $qUM_\textrm{BH}=10^{-3}$. This is an approximation to the Schwarzschild limit. We observe that the comparison of the resulting amplification factors (shown in dashed green in Fig. \ref{fig:test_Schwarzschild}) from both codes indicate a rather sharp agreement. In fact, the Schwarzschild limit is achieved for a variety of small compactnesses, as long as the BH charge is rather small as shown in Fig. \ref{fig:test_Schwarzschild}.

%%%%%%%%%%%%%%%%%%%%%%%%%%%%%%%%%%%%%%%%%%
\subsection{Massless charged scalar waves}
%%%%%%%%%%%%%%%%%%%%%%%%%%%%%%%%%%%%%%%%%%

%%%%%%%%%%%%%%%%%%%%%%%%%%%%%%%%%%%%%%%%%%%%%%%%%%%
\subsubsection{Comparison of amplification factors: non-vacuum BHs versus RN BHs}
%%%%%%%%%%%%%%%%%%%%%%%%%%%%%%%%%%%%%%%%%%%%%%%%%%%

\begin{figure}
     \includegraphics[width=0.45\textwidth]{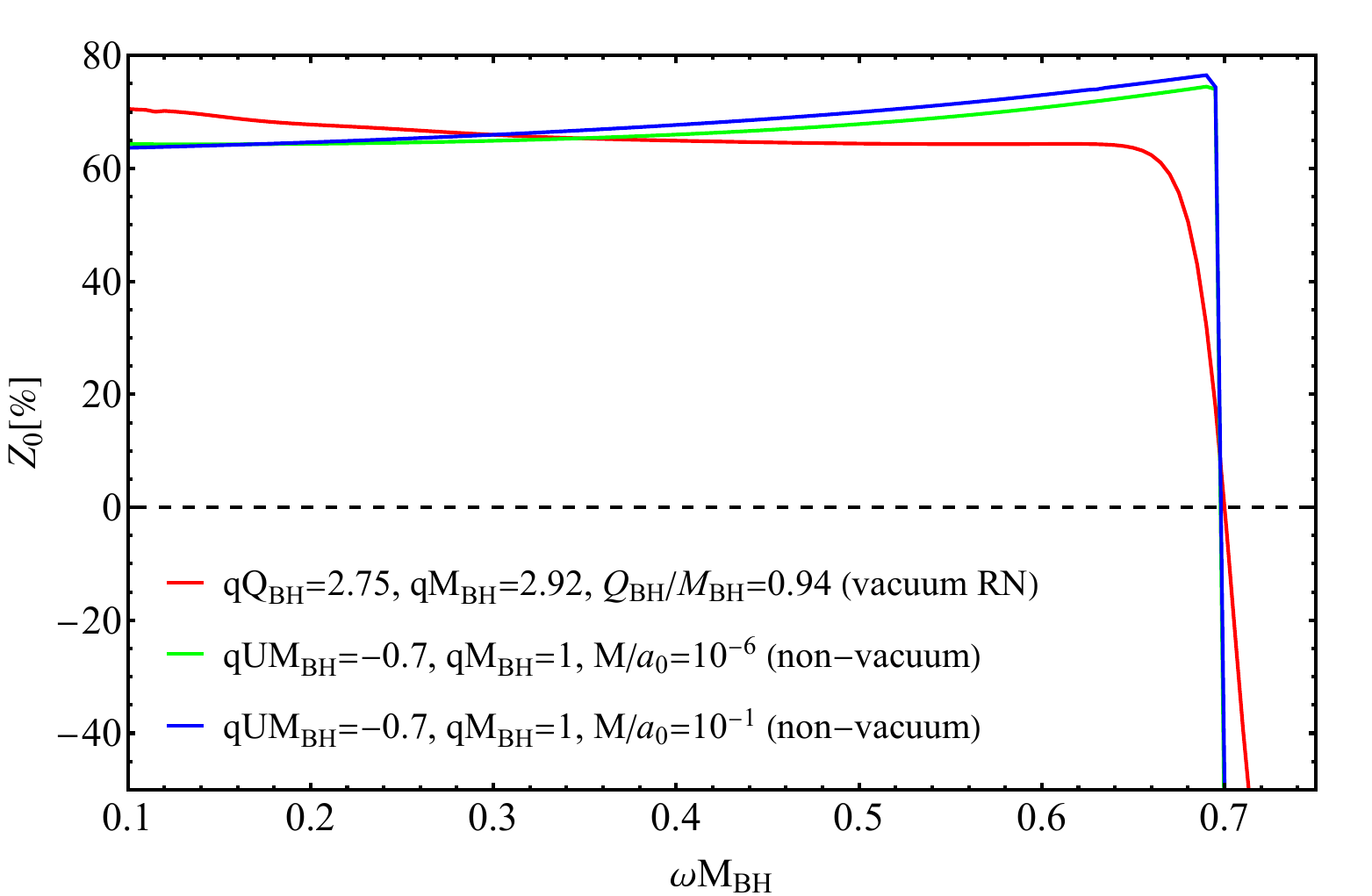}
     \caption{Amplification factors of a charged massless scalar field scattering off a RN BH (red curve) and the BH described by the Eq. \eqref{eq:Stelea_line_element} with two difference choices of compactness (green and blue curves). In all cases, the scalar field charge is equal to unity.}
     \label{fig:RN_vs_Stelea}
\end{figure}

As discussed above, the spacetime involved in this analysis contain a charged BH centralized at a Hernquist dark matter halo. In the limit where the halo disappears ($M=a_0 = 0$) the central object asymptotes to a RN BH (after a specified coordinate transformation). In Fig. \ref{fig:RN_vs_Stelea} we compare the amplification factors of massless charged scalar fields scattering off the BH in Eq. \eqref{eq:Stelea_line_element} to the  amplification factors of a RN BHs impinged by charged scalar fields with the same scalar charge. The charge coupling for our configuration is set to $qUM_\textrm{BH}=-0.7$, though the results for other charge couplings are qualitatively similar. At the same time, the vacuum RN amplification factors of charged scalar fields are calculated for a particular combination of $M_\textrm{BH}$ and $Q_\textrm{BH}$ (see Eqs. \eqref{eq:ADM_and_charge}) so that the resulting (red) curve lies as close to those of the configuration \eqref{eq:Stelea_line_element} as possible\footnote{There is always some error involved between the mass and charge of the RN and those of Eq. \eqref{eq:Stelea_line_element} due to the fact that for astrophysical configurations the halo contributes to the $\textrm{ADM}$ mass and the asymptotic charge of the system, despite the compactness being small.}.

The amplification factors shown in Fig. \ref{fig:RN_vs_Stelea} depict the aforementioned discussion. The red curve shows the amplification factors of charged scalar waves scattering off a RN BH with appropriate chosen BH mass and charge. The corresponding amplification factors resulting from the configuration \eqref{eq:Stelea_line_element} are shown with blue and green, where different compactnesses have been considered in order to understand how superradiance is altered when the halo compactness varies (an in depth discussion will be presented in Sec. \ref{full_analysis_massless}). In general, we observe a similar behavior between the amplification factors for the RN BH and the system in study, since superradiance ends at the same frequency (this is connected with the effective potential Eq. \eqref{eq:effective_potential} remaining almost unaffected with the increment of $M/a_0$). The deviations observed at the small and large superradiant frequency regimes are $\sim 15\%$, though for smaller $qUM_\textrm{BH}$ (but significantly greater than the Schwarzschild limit) the deviation drops to $\sim 5\%$. Thus, only superradiant amplifications of massless charged scalars scattering off BHs close to extremality may present more significant deviations, in order for the astrophysical environment to be implicitly `visible'.

Another observation in the curves of Fig. \ref{fig:RN_vs_Stelea} is the steepness appearing in the amplification factors when the high-frequency superradiance boundary is met. Both the RN and the system \eqref{eq:Stelea_line_element} present a sharper, than usual, transition to non-superradiant frequencies of scalar waves, together with significant amplification $Z_0$. This occurs due to the value of the BH and ADM masses involved in the calculation of superradiant frequencies. Indeed, even when $M_\textrm{BH}=1$ in Eq. \eqref{eq:Stelea_line_element}, the ADM mass of the configuration is larger than unity (see Eq. \eqref{eq:ADM_mass}), which results to steep boundaries of superradiant frequency ranges. In App. \ref{App:A} we discuss in depth how the same effects can take place in vacuum RN BHs by simply increasing their mass and charge (see Fig. \ref{fig:RN_vary_M_l=0}). 

%%%%%%%%%%%%%%%%%%%%%%%%%%%%%%%%%%%%%%%%%%%%%%%%%%%%
\subsubsection{Parameter space analysis}\label{full_analysis_massless}
%%%%%%%%%%%%%%%%%%%%%%%%%%%%%%%%%%%%%%%%%%%%%%%%%%%%

\begin{figure}[t]
     \includegraphics[width=0.45\textwidth]{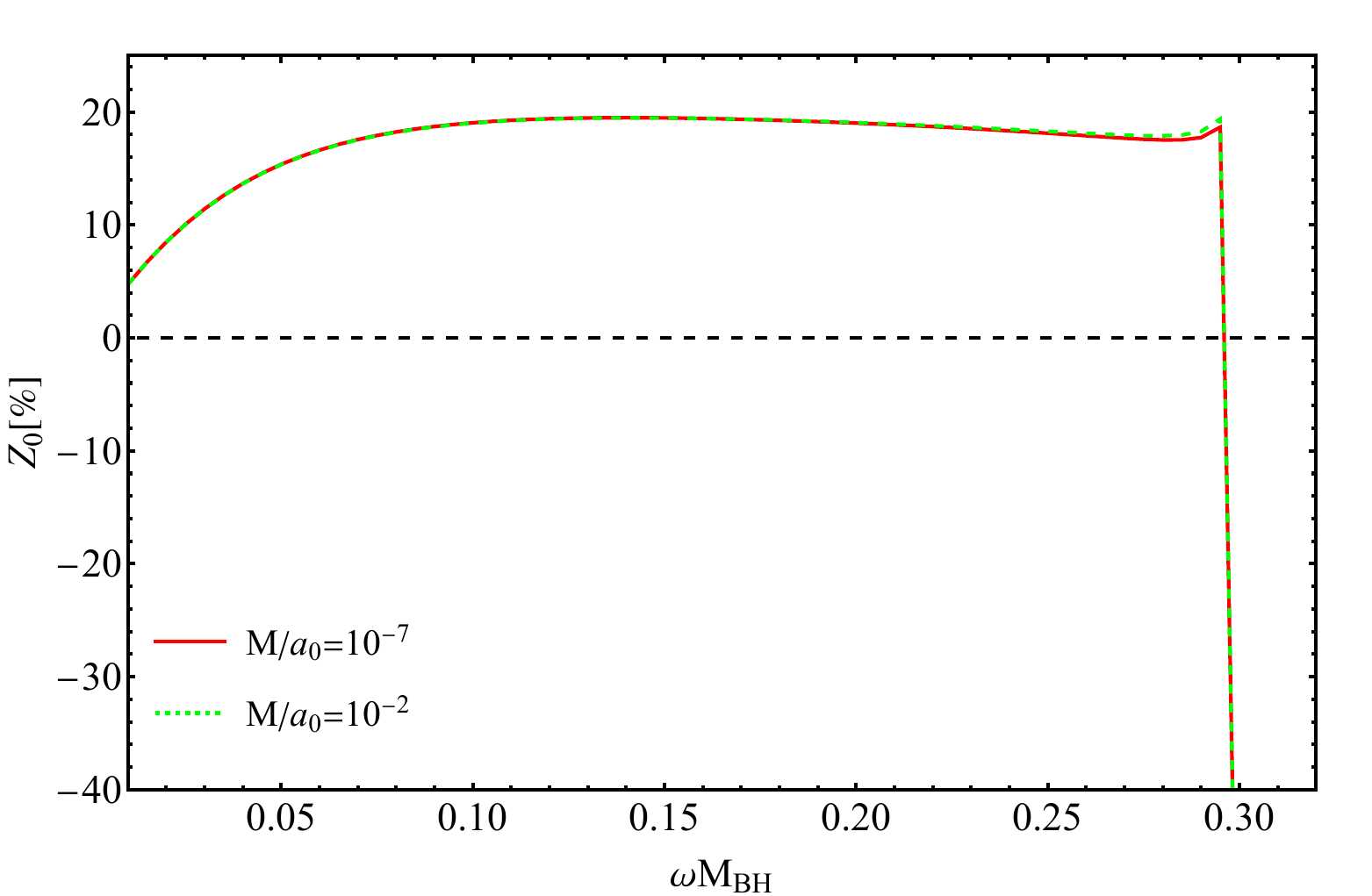} \caption{Amplification factors for $\ell=0$ scalar waves scattering off the spacetime \eqref{eq:Stelea_line_element} with varying compactness $M/a_0$ and charge coupling $qUM_\textrm{BH}=-0.3$.}\label{fig:M/a_0_l=0}
\end{figure}

We commence our examination for massless charged scalar waves, by setting the charge coupling $qUM_{\textrm{BH}}=-0.3$ and varying the compactness of the configuration Eq. \eqref{eq:Stelea_line_element}. By varying the compactness from $M/a_0=10^{-7}$, that is practically a vacuum BH, to $M/a_0=10^{-2}$ we observe a minuscule percentage difference that is smaller than $0.1\%$ (see Fig. \ref{fig:M/a_0_l=0}). This is explained from the effective potential in Fig. \ref{fig:potentials}, left panel, which does not seem to change drastically when the compactness of the halo is increased.

\begin{figure}[t]
     \includegraphics[width=0.45\textwidth]{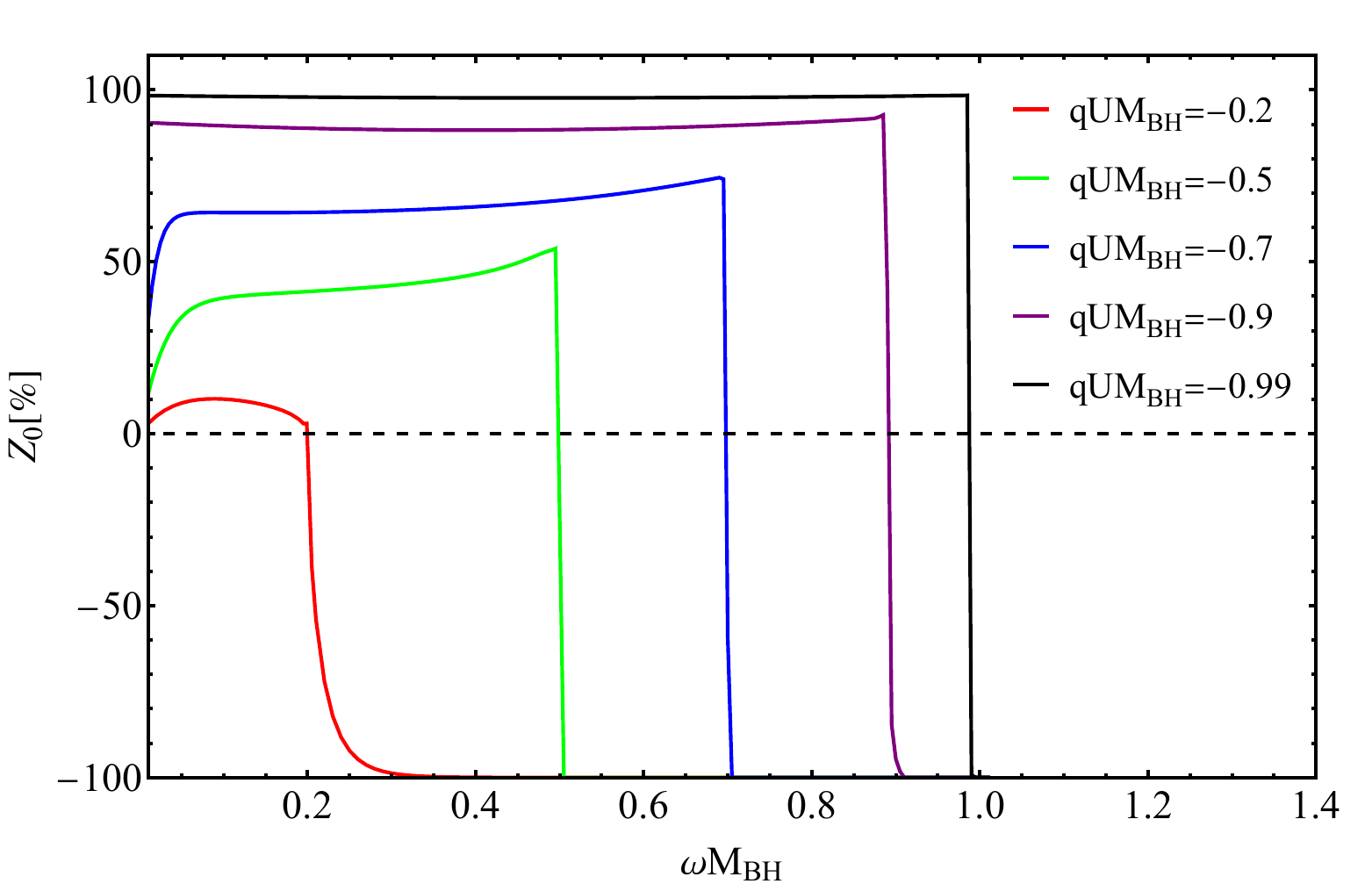} \caption{Amplification factors of $\ell=0$ charged scalar fields with compactness $M/a_0=10^{-4}$, scalar charge $qM_\textrm{BH}=1$ and varying charge coupling $qUM_\textrm{BH}$.}\label{fig:qU_l=0}
\end{figure}

\begin{figure}[t]
     \includegraphics[width=0.45\textwidth]{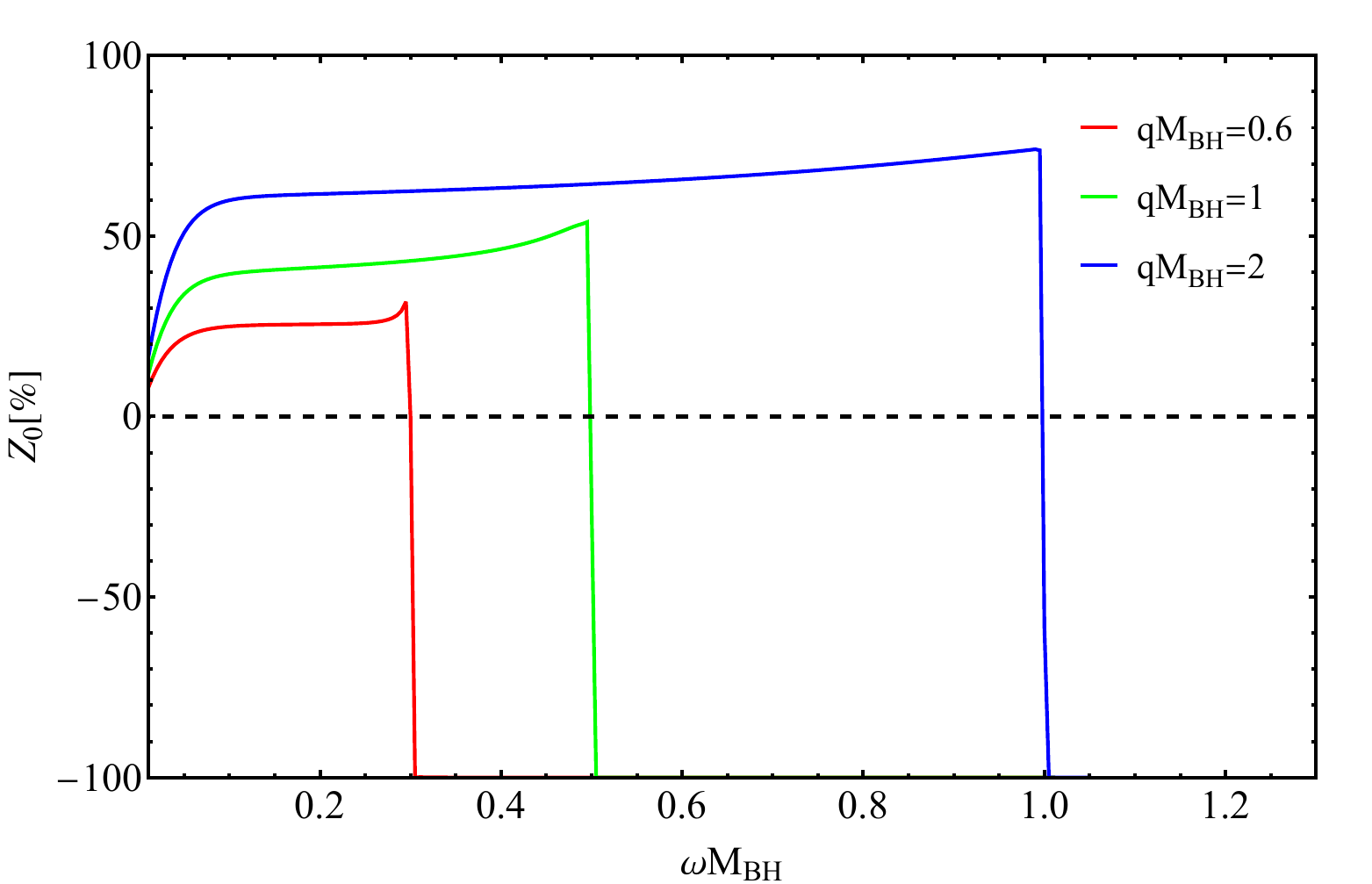} \caption{Amplification factors of $\ell=0$ charged scalar fields with compactness $M/a_0=10^{-4}$, charge coupling $qUM_\textrm{BH}=-0.5$ and varying scalar charge $qM_\textrm{BH}$.}\label{fig:q_l=0}
\end{figure}

\begin{figure}[t]
     \includegraphics[width=0.45\textwidth]{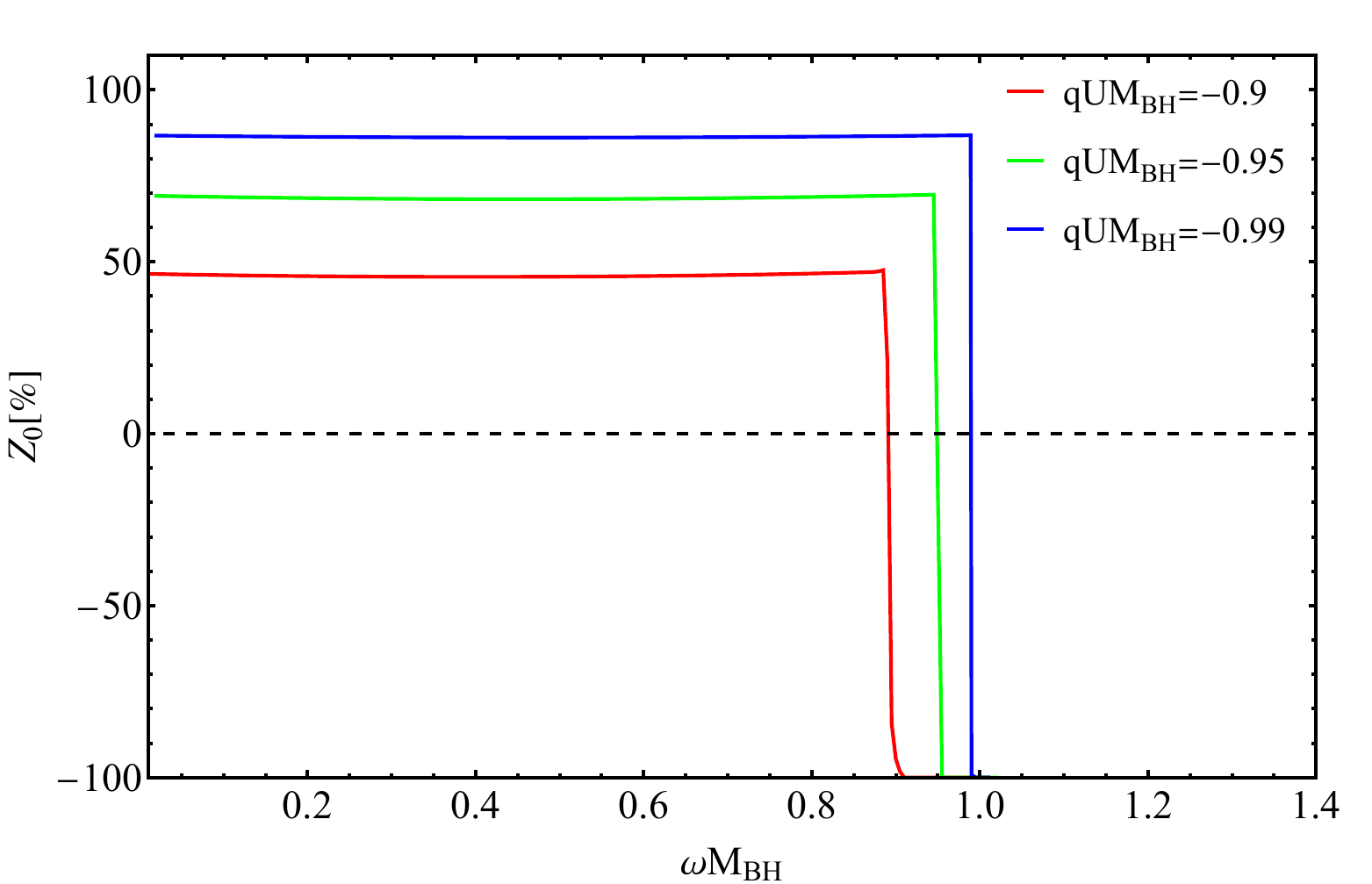} \caption{Amplification factors of $\ell=1$ charged scalar fields with compactness $M/a_0=10^{-4}$, scalar charge $qM_\textrm{BH}=1$ and varying charge coupling $qUM_\textrm{BH}$.}\label{fig:qU_l=1}
\end{figure}

Fixing the compactness to $M/a_0=10^{-4}$, we show in Fig. \ref{fig:qU_l=0} that the increment of the charge coupling $qUM_\textrm{BH}$ leads to higher amplification factors for $\ell=0$ charged massless scalar waves. This also occurs in vacuum RN BHs, i.e. the amplification factors for $\ell=0$ charged scalar waves are steadily increasing as $qUM_\textrm{BH}$ becomes larger. Different combinations of $qM_\textrm{BH}$ and $qUM_\textrm{BH}$ can also lead to significant amplification. From Fig. \ref{fig:qU_l=0}, we can further perform a convergence test, by matching the $\omega M_\textrm{BH}$ where $Z_0=0\%$ with the one predicted by Eq. \eqref{eq:Stelea_superradiant_condition}. As expected the predicted frequency where superradiance ends matches perfectly that of the numerical results presented in Fig. \ref{fig:qU_l=0}.

On the other hand, by keeping the halo compactness $M/a_0=10^{-4}$ and the BH charge $U=-0.5$ constant we vary the scalar charge in order to examine the effect of both charges to the superradiant amplification factors. We observe that varying the scalar field's charge $q M_\textrm{BH}$ leads to higher amplification factors. Figure \ref{fig:q_l=0} shows that the increment of the scalar charge affects the amplification factors significantly (as long as $U<1$). However, by examining the black and blue curves in Fig. \ref{fig:qU_l=0} and Fig. \ref{fig:q_l=0}, respectively, we observe that for a given value of $qUM_\textrm{BH}$, increasing $U$ is more effective in enhancing superradiance than increasing $qM_\textrm{BH}$. This occurs because the increment of $U$ leads to a more efficient way of enhancing the electromagnetic coupling $qQ_{\textrm{BH}}$ than increasing $q$ (see relationship between $U$ and $Q_{\textrm{BH}}$ in Eq. (\ref{eq:BH_charge})).
In the literature, for RN we do not observe large amplification factors as those in Fig. \ref{fig:qU_l=0} (see for instance Fig. (11) of \cite{Brito:2015oca}). However, this should not be interpreted as an effect of the environment amplifying superradiance. Indeed, it is common to assume in numerical codes that $M_{\textrm{BH}}=1$ and choose the BH charge so that $Q_{\textrm{BH}} < M_{\textrm{BH}}$. Here, through an appropriate choice of $U$, we can achieve larger BH charges (without reaching extremality), resulting in larger values for the coupling $q Q_{\textrm{BH}}$ and consequently larger amplification factors. Similarly, for a RN black hole one can obtain larger charges by choosing $M_{\textrm{BH}} > 1$, and as a result a larger amplification with a steeper superradiant boundary (see Fig. \ref{fig:RN_vs_Stelea}).

Again, Fig. \ref{fig:q_l=0} proves that the numerical frequency for which $Z_\ell=0\%$ is very close to the predicted one from Eq. \eqref{eq:Stelea_superradiant_condition}. For completeness, we also calculate the amplification factors for $\ell=1$ charged massless scalar waves. The results are shown in Fig. \ref{fig:qU_l=1}, where the charge coupling is $\sim 1$. We expect that the same will occur for $\ell> 1$ combined with larger charge couplings. Therefore, we conclude that superradiance takes place in the BHs under investigation and can produce amplification factors that are, at least, similar with respect to those in a vacuum RN BH.

%%%%%%%%%%%%%%%%%%%%%%%%%%%%%%%%%%%%%%%%%
\subsection{Massive charged scalar waves}
%%%%%%%%%%%%%%%%%%%%%%%%%%%%%%%%%%%%%%%%%

\begin{figure}[t]
     \includegraphics[width=0.45\textwidth]{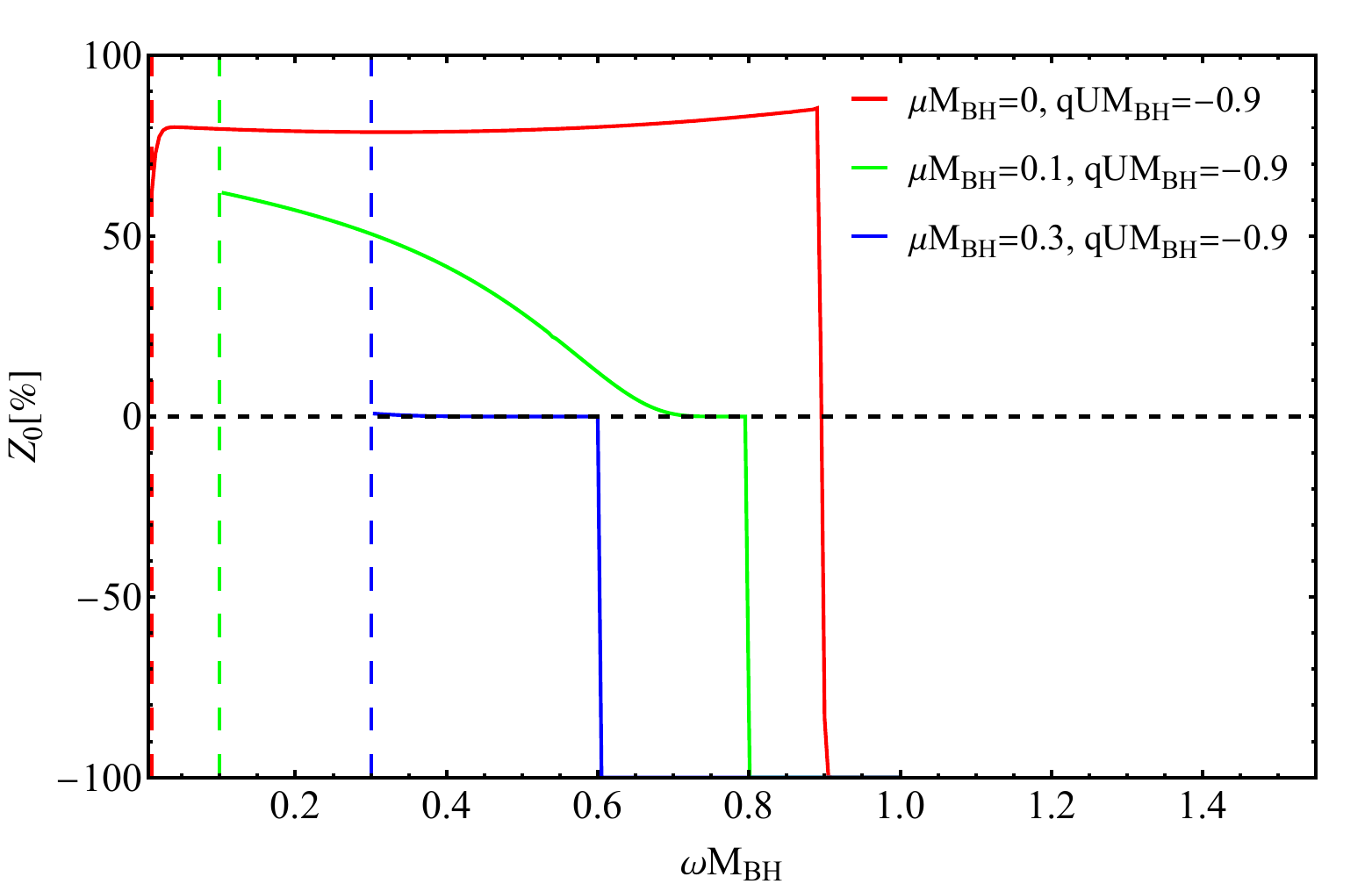} \caption{Amplification factors of $\ell=0$ charged massive scalar fields with compactness $M/a_0=10^{-4}$, charge coupling $qUM_\textrm{BH}=-0.9$, scalar charge $qM_\textrm{BH}=1.125$ and varying scalar mass $\mu M_\textrm{BH}$. The vertical dashed lines correspond to the lower limit above which superradiance takes place.}\label{fig:mu_l=0}
\end{figure}

\begin{figure}[t]
     \includegraphics[width=0.45\textwidth]{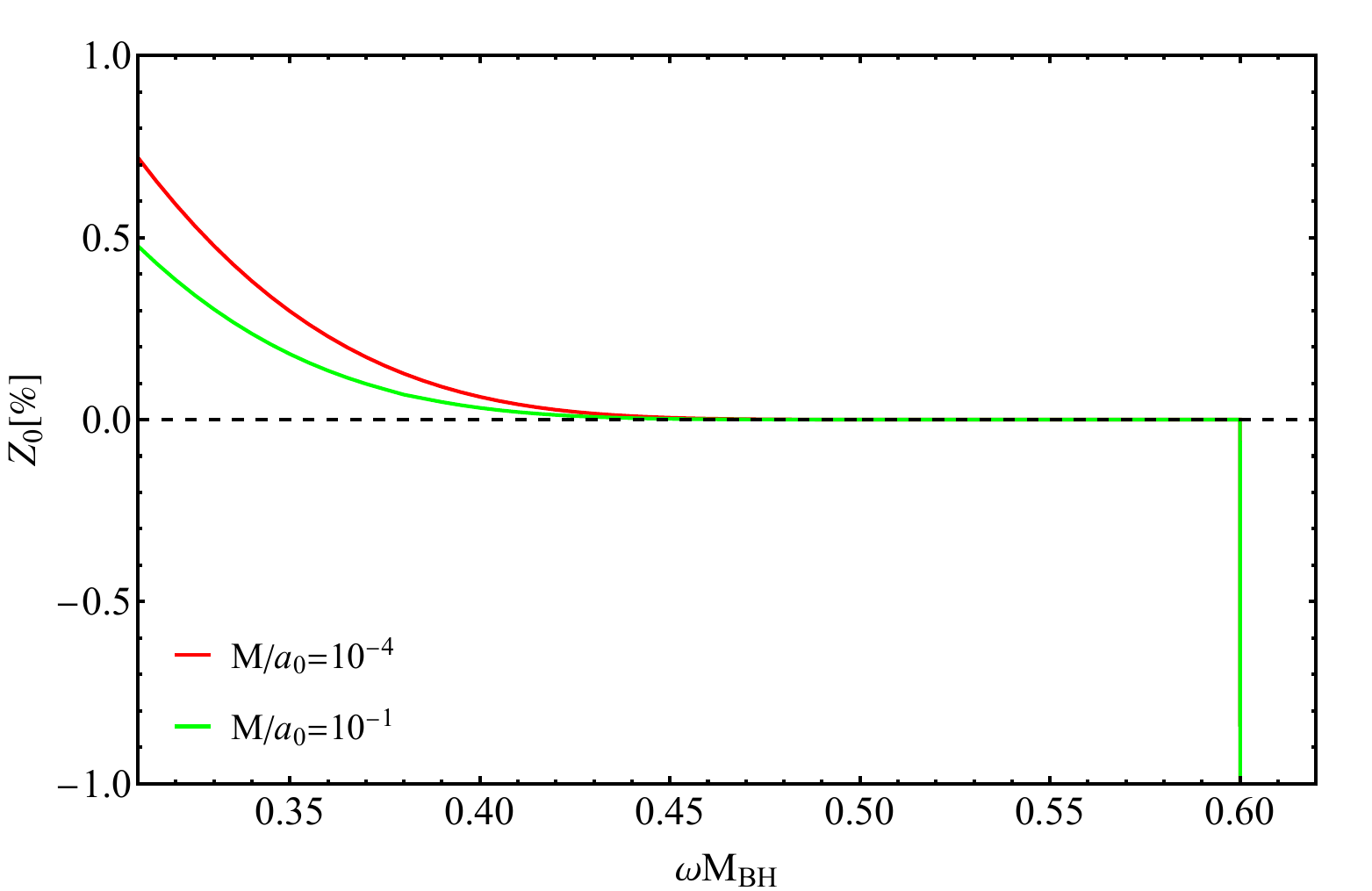} \caption{Amplification factors of $\ell=0$ charged massive scalar fields with charge coupling $qUM_\textrm{BH}=-0.9$, scalar charge $qM_\textrm{BH}=1.125$, scalar mass $\mu M_\textrm{BH}=0.3$ and varying compactness $M/a_0$.}\label{fig:mu_varycomp_l=0}
\end{figure}

The cases of massive charged scalar waves impinging the BH in Eq. \eqref{eq:Stelea_line_element} are presented in Figs. \ref{fig:mu_l=0} and \ref{fig:mu_varycomp_l=0}. The addition of mass on the scalar field is examined in Fig. \ref{fig:mu_l=0}. We observe that the mass serves as a friction mechanism, i.e. as the mass of the scalar wave is increased the superradiant range is decreasing (in accord with Eq. \eqref{eq:superradiant_R_Stelea_massive}) so that superradiance begins at $\mu$ and ends at $-\mu+|qU|$. 

At the same time, the amplification factors are decreasing with the increment of $\mu M_\textrm{BH}$; a sign of friction behavior that can be explained from the effective potential in Fig. \ref{fig:potentials}, right panel, where the addition of mass elevates the effective potential to asymptotically positive values. This means that if $\omega<\mu$ then there are no propagating waves but rather decaying solutions that never reach the effective potential. This is also the reason why the superradiant relation in Eq. \eqref{eq:superradiant_R_Stelea_massive} has both a positive minimum and a maximum value for superradiance to occur. In fact, Fig. \ref{fig:mu_l=0} shows, besides the amplifications factors for each case, vertical dashed lines that correspond to the minima of $\omega M_\textrm{BH}$, above which superradiance occurs.

It is important to note that the increment of compactness affects the amplification factors of massive, charged scalar fields significantly. In particular, for the cases shown in Fig. \ref{fig:mu_varycomp_l=0}, where $\mu M_\textrm{BH}=0.3$ and $qUM_\textrm{BH}$, the percentage difference between the two curves, corresponding to two different compactnessess, in the low frequency regime is $50\%$. This is due to the gravitational interaction between the mass of the scalar field and the dark matter halo mass that can significantly change when the BH is close to extremality and therefore the gravitational and electrogmanetic interactions clash. This means that massive, highly charged scalar fields that scatter off non-vacuum, near extremally-charged BHs with large halo compactness can show significant deviations in the amplification factors with respect to those occurring in vacuum RN BHs. Even though these compact configurations are cannot describe the distribution of dark matter around galaxy, they can simulate dark matter overdensities and, in general, BH hair.

%%%%%%%%%%%%%%%%%%%%%%%%%%%%
\section{Concluding remarks}
%%%%%%%%%%%%%%%%%%%%%%%%%%%%

Supermassive BHs typically reside in galactic cores where astrophysical environments, such as accretion disks and large-scale dark matter halos, affect the evolution and propagation of fields due to environmental effects. Therefore, the study of non-vacuum BHs that include astrophysical environments or generic hair formations around them is currently fundamental for GW astronomy. More precisely, new exact and fully-relativistic solutions to the fields equations have recently been obtained in order to study various effects. Here, we have focused on the phenomenon of superradiance on an exact charged BH solution embedded in a Hernquist dark matter halo distribution. By scattering massless and massive charged scalar fields on such configuration, we observed through the amplification factors how the dark matter environment affects superradiance.

We have studied the scattering of massless and massive scalar monochromatic waves with variant frequency $\omega M_\textrm{BH}$, in order to assess how the parameter space affects the superradiant amplification of those waves. We first found the perturbation equation of a charged BH, centered at an, otherwise generic, dark matter halo and with appropriate boundary conditions we performed a scattering experiment in order to calculate the amplification factors of charged scalar waves that scatter off such spacetime. The newly-obtained radial equation of motion contains a charge coupling term at the effective potential, between the BH and scalar charge, and further possesses a first-order derivative term that breaks the Schr\"odinger-like form of the linear perturbation equation of motion. This first-order term indicates potential effects of friction between the scattered waves and the dark matter halo.

We showed that the effective potential outlined in Eq. \eqref{eq:effective_potential} has a weak dependence on the halo compactness but on the contrary, a strong dependence on the charge coupling and the scalar mass. In fact the last two parameters seem to compete when a charged massive scalar waves impinges the BH under study. The charge coupling is reducing the asymptotic value of the potential to negative values, thus allowing for deeper regions in $r/M_\textrm{BH}$ where superradiance can be more efficient and span in a wider frequency band. On the other hand, the scalar mass has the opposite effect on BH superradiance, i.e. the larger it gets the more positive the potential becomes asymptotically, and thus the region for bound states is minimized. This leads to an attenuated superradiant amplification, as well as an amplification for a shorter frequency regime, in accord with Eq. \eqref{eq:superradiant_R_Stelea_massive}.

Overall, we conclude that the study of superradiance in such BH configurations depends on the charges in play (BH and scalar), as well as the scalar mass. In fact, we found a significant dependence of the amplification factors on the compactness of the halo, which can be attributed to the subdominant effect of it at the effective potential and the presence of a gravitational interaction between the halo and the scalar mass. In fact, the difference between the two amplification factors in Fig. \ref{fig:mu_varycomp_l=0} shows that a larger scalar mass, or scalar coupling, should, in principle, amplify the difference between the factors of the two cases with different compactness. Nevertheless, the effect, though unphysical for galaxies, can be potentially physical for BH hair and dark matter spikes. For astrophysical values though, superradiance is not enough to disentangle the environmental from the BH effects.

We expect that the study of more environments, and in particular rotating BHs in dark matter halos, will significantly help in distinguishing the existence of BH environments, as well as the difference between such surroundings, with combined analysis of high-energy phenomenology such as superradiance, quasinormal modes and other perturbative calculations, such as extreme-mass-ratio binaries.

%%%%%%%%%%%%%%%%%%%%%%%
\begin{acknowledgments}
The authors would like to thank Paolo Pani for significant help on the superradiance code and on the finalization of this draft. We acknowledge partial support by the MUR PRIN Grant 2020KR4KN2 ``String Theory as a bridge between Gauge Theories and Quantum Gravity'' and by the MUR FARE programme (GW-NEXT, CUP:~B84I20000100001). 
\end{acknowledgments}
%%%%%%%%%%%%%%%%%%%%%%%

%%%%%%%%%
\appendix
%%%%%%%%%

%%%%%%%%%%%%%%%%%%%%%%%%%%%%%%%%%%%%%%%%%%%%%%%%%%%%
\section{Reissner-Nordstr\"om BH superradiance and amplification factor steepness}\label{App:A}
%%%%%%%%%%%%%%%%%%%%%%%%%%%%%%%%%%%%%%%%%%%%%%%%%%%%

In this section we present some examples of amplification factors of charged massless scalar waves scattering off RN BHs. The purpose of this presentation is to better understand the steepness of the amplification factors of the metric \eqref{eq:Stelea_line_element}, together with the rather high amplification factors appearing than naively can be explained by the existence of the dark matter halo.

\begin{figure}[h]
     \includegraphics[width=0.45\textwidth]{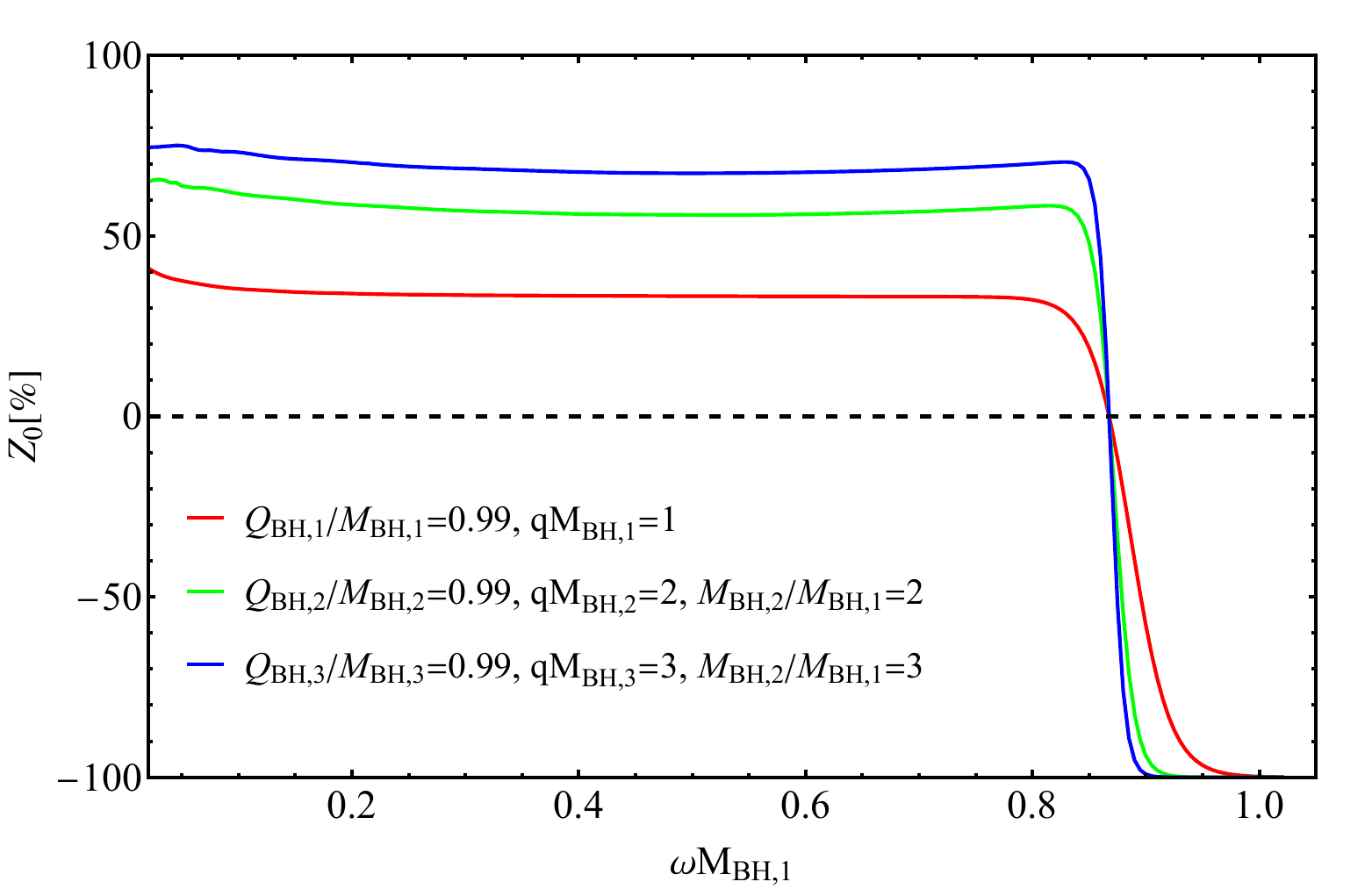} \caption{Amplification factors of $\ell=0$ charged massless scalar fields scattering off a RN BH with scalar charge $qM_\textrm{BH,1}=1$ and varying BH charge/mass.}\label{fig:RN_vary_M_l=0}
\end{figure}

Figs. \ref{fig:M/a_0_l=0} - \ref{fig:mu_varycomp_l=0} depict a large variety of amplification factors, due to superradiance. A patterns that is repeatedly observed is a quite steep end of superradiance at particular frequencies. Nonetheless, the end of superradiance, i.e. $Z_\ell=0$, always satisfies the superradiant relation \eqref{eq:massive_superradiant_condition}. Despite the fact that RN BHs present smoother transitions to non-superradiant monochromatic wave frequencies they are usually studied by choosing geometrized units and subsequently setting $M_\textrm{BH}=1$, which effectively changes length dimensionality to dimensionless.

A reasoning behind the sharp saturation of superradiance in configurations like that described by \eqref{eq:Stelea_line_element} is the fact that the total (ADM) mass does not equal $M_\textrm{BH}$ but rather is equal to Eq. \eqref{eq:ADM_mass}, which is not necessarily equal to unity. This introduces a new scale in the calculation of the amplification factors that is responsible for their steepness observed.

To better understand how $M_\textrm{BH}\neq M_\textrm{ADM}$ affects the amplification factors and superradiance of charged scalars on \eqref{eq:Stelea_line_element} we calculate $Z_0$ for RN BHs for which the mass is increased from unity to $3$. As observed in Fig. \ref{fig:RN_vary_M_l=0}, increasing the RN BH mass leads to amplification factors that, first, are larger with respect to that where $M_\textrm{BH}=1$ and in particular, the steepness close to the end of superradiance increases rapidly. The fact that Figs. \ref{fig:M/a_0_l=0} - \ref{fig:mu_varycomp_l=0} are not scaled with respect to $M_\textrm{ADM}$ but rather with $M_\textrm{BH}$ leads to the increment of steepness, which is also quite obvious in Fig. \ref{fig:RN_vary_M_l=0}, for a rather small increment of the BH mass. Since, whatever the compactness $M/a_0$, $M_\textrm{ADM}\gg M_\textrm{BH}$ for realistic astrophysical scenarios, the steepness observed in the main text is natural and occurs also in vacuum RN BHs. At the same time, increasing the mass of the BH leads to higher amplification factors, therefore this also explains the rather large factors appearing in Figs. \ref{fig:M/a_0_l=0} - \ref{fig:mu_varycomp_l=0}. Thus, the astrophysical environment surrounding the charged BH in Eq. \eqref{eq:Stelea_line_element} plays an important role in the superradiant amplification of charged scalar fields, only when the field has a mass term in order to gravitationally interact with the dark matter halo.

%%%%%%%%%%%%%%%%%%%%%
\bibliography{biblio}
%%%%%%%%%%%%%%%%%%%%%

\end{document}